\documentclass[pdflatex,sn-mathphys-ay]{sn-jnl}

\usepackage{graphicx}%
\usepackage{multirow}%
\usepackage{amsmath,amssymb,amsfonts}%
\usepackage{amsthm}%
\usepackage{mathrsfs}%
\usepackage[title]{appendix}%
\usepackage{xcolor}%
\usepackage{textcomp}%
\usepackage{manyfoot}%
\usepackage{booktabs}%
\usepackage{algorithm}%
\usepackage{algorithmicx}%
\usepackage{algpseudocode}%
\usepackage{listings}%

\usepackage{graphicx}%
\usepackage{multirow}%
\usepackage{amsmath,amssymb,amsfonts}%
\usepackage{amsthm}%
\usepackage{mathrsfs}%
\usepackage[title]{appendix}%
\usepackage{xcolor}%
\usepackage{textcomp}%
\usepackage{manyfoot}%
\usepackage{booktabs}%
\usepackage{algorithm}%
\usepackage{algorithmicx}%
\usepackage{algpseudocode}%
\usepackage{listings}%
\usepackage{soul}%
\usepackage{subcaption}%
\usepackage{booktabs}
\usepackage{makecell}
\usepackage[utf8x]{inputenc}

\theoremstyle{thmstyleone}%
\theoremstyle{thmstyletwo}%

\theoremstyle{thmstylethree}%

\begin{document}

\title[Article Title]{Bayesian Optimization of Genetic Algorithm Hyperparameters in a Multi-Fidelity Framework for Efficient Lattice Material Design}


\author[1,2]{\fnm{Sergei} \sur{Zorkaltsev}}\email{zorkaltsevsergey@gmail.com, 0009-0005-9731-1829}

\author[1]{\fnm{Maciej} \sur{Haranczyk}}\email{maciej.haranczyk@imdeamaterials.org, 0000-0001-7146-9568}

\author*[1]{\fnm{Christina} \sur{Schenk}}\email{christina.schenk@imdea.org, 0000-0002-7817-6757}

\affil[1]{\orgname{IMDEA Materials Institute}, \orgaddress{\street{Eric Kandel 2}, \city{Getafe}, \postcode{28906}, \state{Madrid}, \country{Spain}}}

\affil[2]{\orgdiv{Department of Materials Science and Engineering and Chemical Engineering}, \orgname{Universidad Carlos III de Madrid}, \orgaddress{\street{Av. Universidad 90}, \city{Leganes}, \postcode{28911}, \state{Madrid}, \country{Spain}}}

\abstract{This study presents a multi-fidelity framework for the systematic optimization of genetic algorithm (GA) hyperparameters. The framework integrates three fidelity levels: high-fidelity Fast Fourier Transform (FFT) homogenization for validation, a medium-fidelity 3D convolutional neural network surrogate for rapid property evaluation, and a low-fidelity Gaussian process (GP) surrogate within a Bayesian optimization (BO) framework to guide the hyperparameter search. Various acquisition functions are evaluated, with logNEI achieving the best performance by effectively accounting for the noise inherent in GA evaluations. The proposed framework identifies hyperparameter configurations that enable a 25-generation GA run to achieve elastic modulus values comparable to those obtained in a full 75-generation optimization. Furthermore, introducing a penalized BO objective significantly reduces the number of required lattices with only minor decreases in absolute achieved elastic modulus, revealing a practical trade-off between performance and the number of structures that must be evaluated. High-fidelity FFT validation verifies the effectiveness of the surrogate-driven optimization strategy. The optimized hyperparameters allow for rapid convergence, eliminate the need for lattice mutation, and reduce the overall computational cost by 24\% (from 225 to 171 hours) while preserving mechanical performance. These results demonstrate the potential of multi-fidelity optimization as an efficient and practical approach for GA hyperparameter tuning and future experimental lattice design studies.}

\keywords{Bayesian optimization, genetic algorithm, multi-fidelity optimization, convolutional neural network, lattice materials, hyperparameter optimization}



\maketitle

\section{Introduction}

Lattice materials are architectured structures formed by periodic or aperiodic arrangements of strut-based cells, and are attractive because their geometry can be used to tailor properties such as energy absorption \citep{energy_abs1, energy_abs2, energy_abs3}, heat and fluid transport \citep{transfer1, transfer2,transfer3}, elastic response \citep{mechanical1, mechanical2}, and low-density structural performance \citep{lightweight1, lightweight2}. Combined with the rapid development of additive-manufacturing routes such as selective laser melting \citep{slm}, electron beam melting \citep{ebm}, fused deposition modeling \citep{fdm}, and stereolithography \citep{sla}, this has made lattice design a central problem in lightweight engineering, biomedical devices, aerospace, and automotive applications. 

To address this problem, the field has explored a wide range of computational design strategies. These include topology optimization of individual unit cells \citep{spear_computational}, multiscale optimization of cell topology \citep{multiscale_optim}, and multi-topology lattice design based on libraries of precomputed cells \citep{rapid_modeling}. More recently, machine-learning and inverse-design approaches have been introduced to reduce the cost of evaluating candidate architectures, including neural-network surrogates for mechanical prediction \citep{bcc_nn}, generative models for unit-cell design \citep{nn_generation, inverse_gans}, and hybrid frameworks that couple data-driven models to physics-based simulations \citep{dos2022inverse, dos2024deep, wang2025inverse, xiang2025decoupled}. In parallel, evolutionary algorithms have been used to navigate large, discrete, and multi-modal design spaces, either as standalone optimizers or in combination with homogenization, finite-element methods, and surrogate models \citep{topology_ga, sinusoidal_ga_ml, pragmatic_design}.

Despite these advances, two challenges remain central. First, the combinatorial complexity of heterogeneous lattice systems causes the design space to grow rapidly as multiple unit-cell types and spatial arrangements are considered simultaneously. Second, accurate evaluation of candidate structures remains computationally expensive, particularly when high-fidelity homogenization or finite-element simulations are required. As a result, practical optimization workflows increasingly rely on multi-fidelity strategies that combine fast approximations with selective use of expensive simulations.

In our previous work, we proposed such a multi-fidelity framework for the optimization of high-entropy lattice materials. The approach considered $4 \times 4 \times 4$ lattice assemblies constructed from five predefined unit-cell types and combined two fidelity levels: FFT-based computational homogenization as the high-fidelity evaluator, a 3D convolutional neural-network surrogate as a medium-fidelity predictor connected with a GA as the global search engine. This hierarchical workflow reduced the number of FFT simulations by approximately 90\%, while achieving an increase of up to 91.5\% in the specific elastic modulus compared to pure single-topology lattices. The study also showed that different objective formulations lead to distinct optimized structural patterns.

However, the efficiency of the optimization process itself remained strongly dependent on the choice of GA hyperparameters, including population size, crossover probability, mutation rate, and selection pressure. Since evaluating each hyperparameter configuration requires repeated optimization runs, tuning the GA becomes a computationally expensive black-box optimization problem in its own right. To address this limitation, the present work introduces BO into the workflow as a low-fidelity supervisory layer for automated hyperparameter tuning. As illustrated in Fig.~\ref{summary}, the proposed framework organizes the optimization process across three fidelity levels: FFT simulations provide high-fidelity mechanical evaluations, the CNN surrogate serves as a medium-fidelity model of the lattice properties to accelerate GA search, and BO employs a GP as a low-fidelity model of the GA hyperparameter-response landscape. By coupling these components within a unified multi-fidelity strategy, the objective is not only to improve the final mechanical performance of optimized lattices, but also to increase the efficiency, robustness, and reproducibility of the overall search process.

The main contributions of this work are threefold. First, a three-level multi-fidelity framework combining FFT homogenization, a CNN surrogate model, and GP-based BO is developed for automated GA hyperparameter tuning. Second, several BO acquisition functions are systematically evaluated in the context of noisy GA optimization. Third, a penalized objective formulation is introduced to balance mechanical performance against the number of lattice structures required during optimization.

The remainder of this paper is organized as follows. Section 2 describes the proposed multi-fidelity optimization framework, including the genetic algorithm, FFT homogenization, CNN surrogate model, and BO procedure. Section 3 presents the optimization results, the comparison of the acquisition functions, and the effects of the penalized objective. Finally, Section 4 summarizes the main findings and conclusions.
\section{Methods}\label{methods}

\subsection{General summary}

\begin{figure}[h]
    \centering
    \includegraphics [width=0.85\textwidth]{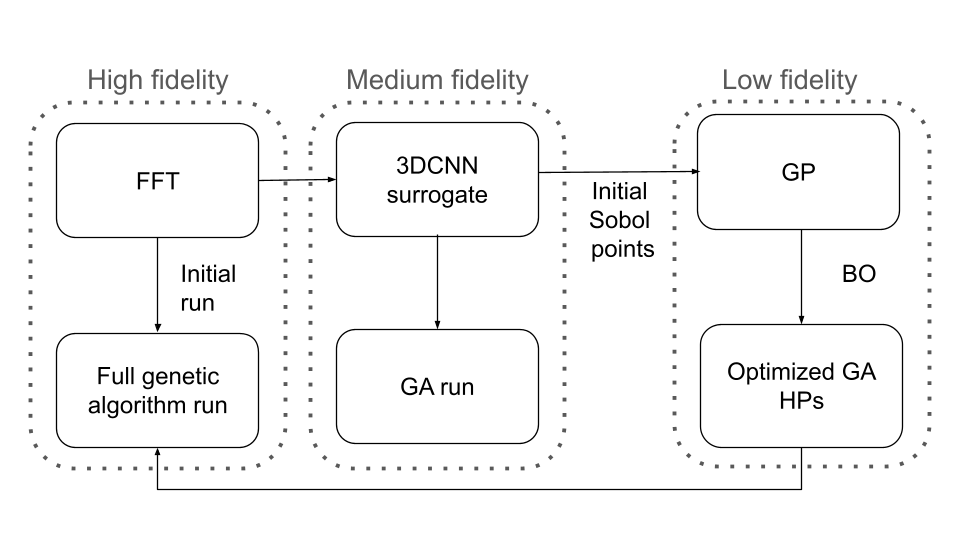}
    \caption{Three-level multi-fidelity optimization framework. FFT homogenization provides high-fidelity mechanical evaluations, a CNN surrogate accelerates lattice-property prediction at the medium-fidelity level, and BO with a GP surrogate guides GA hyperparameter selection at the low-fidelity level. Information flows hierarchically from physical simulations to surrogate-assisted optimization.}\label{summary}
\end{figure}

Figure \ref{summary} illustrates the multi-fidelity workflow developed to optimize the hyperparameters of the GA.
The framework is organized into three fidelity levels. At the low-fidelity level, BO employs a GP model to guide the search for promising hyperparameter configurations. At the medium-fidelity level, a 3D convolutional neural-network (3DCNN) surrogate model accelerates the evaluation of candidate lattice architectures during the GA search. Finally, at the high-fidelity level, FFT-based computational homogenization is used to obtain accurate mechanical-property values for selected structures.

The workflow proceeds hierarchically across these fidelity levels. Initial GA hyperparameter configurations are evaluated through GA optimization runs assisted by the 3DCNN surrogate model at the medium-fidelity level. These initial evaluations, generated from Sobol sampling points, are then used by the BO framework to construct and update the GP surrogate at the low-fidelity level. Based on this information, BO iteratively identifies promising GA hyperparameter configurations with improved optimization performance. Finally, the optimized GA hyperparameters and the corresponding lattice optimization results are validated through high-fidelity FFT simulations.

\begin{figure}[h]
    \centering
    \includegraphics [width=0.85\textwidth]{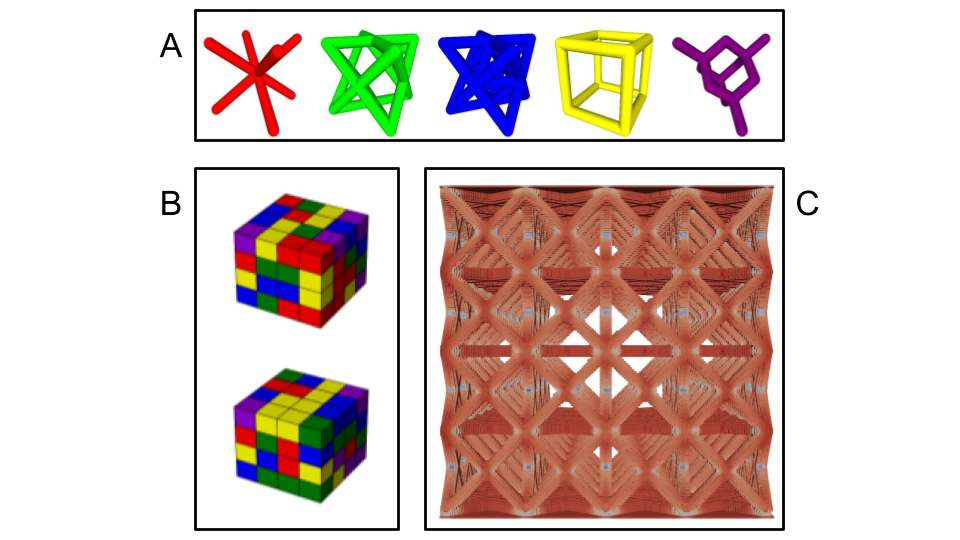}
    \caption{Overview of the lattice representation used throughout the optimization framework. (A) Unit cell types used to construct $4 \times 4 \times 4$ lattices (BCC, FCC, OT, SC, DIA). (B) Lattice encoding used by GA. (C) Voxelized lattice used in FFT simulations. Adapted from \cite{zorkaltsev2026does}.}\label{lattice}
\end{figure}

\subsection{Genetic algorithm}\label{GA}
A genetic algorithm is an optimization approach inspired by natural selection and evolution. In this work, GA is used to optimize the topology of lattice structures by searching for the best-performing arrangement of selected unit cells. 

The genome of each lattice is represented as a $4\times4\times4$ integer matrix, with entries taking values in $[0,4]$, corresponding to the BCC, FCC, OT, SC, and DIA unit cells, respectively (Fig.~\ref{lattice}A). The genome of the lattices can be decoded into a $160^3$ binary matrix, used for high-fidelity simulations (Fig.~\ref{lattice}B,C), by replacing integers with $40^3$ blocks for each cell type. In the case of the binary matrix, void voxels are assigned a value of 0, whereas solid (Inconel-718) voxels are assigned a value of 1. The $4\times4\times4$ representation is used in the genetic operation of GA, while the $160\times160\times160$ binary representation is used in FFT simulations and CNN surrogate model.

The initial population of 1,000 lattice structures was generated by uniform random sampling from [0, 4]. Initial lattices were labelled with the FFT-computed value of specific elastic modulus, calculated for the $z$ principal direction. The term specific elastic modulus, used throughout this manuscript, refers to the elastic modulus value normalized by the lattice's solid fraction. 

Fitness evaluation was performed in two ways. The objective of this GA implementation is to maximize the specific elastic modulus. Most of the candidates were evaluated with a CNN surrogate, trained on the results of the previous GA optimization. For hyperparameter validation, the top-performing structures were also characterized with FFT simulations to fine-tune the surrogate.

Each GA iteration involved several steps. First, the specified number of parents (top-performing lattices) is selected from the previous generation based on their fitness values. The number of selected parents is one of the GA hyperparameters tuned with BO. The parents are randomly shuffled and divided into pairs; if the number is odd, the last lattice is paired with a random one. Each parent pair produces six offspring with a crossover. A uniformly sampled integer $n$ from [1, 3] determines the position of slicing for both parent lattices, and the offspring is constructed by concatenating the first $n$ slices of the first parent with the remaining slices of the second parent. The crossover is repeated for three principal directions. Next, the mutation is applied to the resulting offspring pool. Two hyperparameters control the mutation procedure: $f_{mut}$, which determines the fraction of offspring subjected to mutation, and $f_{cell}$, which determines the fraction of cells modified within each selected offspring. For example, with $f_{mut} = 0.5$, $f_{cell} = 0.1$, and a population of 100 offspring lattices, 50 randomly selected lattices undergo mutation, with 6 out of the 64 cells in each lattice randomly reassigned.

\subsection{Fast Fourier Transform simulations (high fidelity)}\label{sec:fft}
The high-fidelity evaluations in this work were performed using a spectral FFT-based computational solver to obtain the homogenized elastic response of the lattice structures. The FFTMAD code \citep{lucarini2019accuracy} used in this study takes a voxelized 3D lattice as an input, computes the elastic response under specified loading and direction, and returns the elastic modulus together with strain and stress 3D grids.

FFT-based solvers have several advantages compared to traditional finite element methods. First, no meshing is required for FFT - a voxelized phase map can be directly used as an input \citep{MOULINEC199869,schneider2021review}. Secondly, the FFT is computationally more efficient, scaling with \(O(n \log n)\) \citep{cooley1965algorithm,givois2022qftbasedhomogenization,schneider2021review}. The FFTMAD code has been previously tested and validated on lattice materials \citep{lucarini2022adaptation} and compared to the performance of commercially available FEM software \citep{lucarini2022adaptation, zorkaltsev2026does}.

The high-fidelity evaluations in this work were performed using a spectral FFT-based computational solver to obtain the homogenized elastic response of the lattice structures. The FFTMAD code \citep{lucarini2019accuracy} used in this study takes a voxelized 3D lattice as an input, computes the elastic response under specified loading and direction, and returns the elastic modulus together with strain and stress 3D grids.

FFT-based solvers have several advantages compared to traditional finite element methods. First, no meshing is required for FFT - a voxelized phase map can be directly used as an input \citep{MOULINEC199869,schneider2021review}. Secondly, the FFT is computationally more efficient, scaling with \(O(n \log n)\) \citep{cooley1965algorithm,givois2022qftbasedhomogenization,schneider2021review}. The FFTMAD code has been previously tested and validated on lattice materials \citep{lucarini2022adaptation} and compared to the performance of commercially available FEM software \citep{lucarini2022adaptation, zorkaltsev2026does}.

During the FFT simulations, periodicity was maintained only along the loading (z) direction, while the periodic boundary conditions in the transverse (x and y) directions were intentionally broken by introducing a two-voxel layer of void material. This setup reproduces the boundary conditions of a uniaxial compression test and is consistent with the simulation protocol adopted in our previous work \citep{zorkaltsev2026does}.

\subsection{3D convolutional neural-network surrogate (medium fidelity)}\label{sec:cnn}

At the medium-fidelity level, the specific elastic modulus values are predicted by a 3DCNN surrogate model. The network architecture is based on DenseNet \citep{densenet}. It takes a voxelized $160^3$ binary matrix, representing the lattice structure, and outputs the corresponding elastic response value.

The network consists of dense blocks with 6, 12, 24, and 16 layers, respectively, with a growth rate of 32. Dense blocks are connected by transition layers, with $1\times1\times1$ convolution to reduce the number of feature maps, followed by $2\times2\times2$ average pooling. After the final block, batch normalization and adaptive average pooling reduce the feature map to $1\times1\times1$. This feature map is flattened and passed to a fully connected layer to produce a specific elastic modulus prediction. 

Detailed descriptions of performance tests of the architecture and network variations are reported in our previous work \citep{zorkaltsev2026does}.

\subsection{Bayesian optimization}
Together, the FFT solver cf.~Section~\ref{sec:fft}, CNN surrogate cf.~Section~\ref{sec:cnn}, and the GP surrogate introduced below constitute the three fidelity levels of the proposed optimization framework, with each level accelerating the subsequent one.
We consider the problem of maximizing a black-box objective function
\begin{equation}
f:\mathcal{X}\subset\mathbb{R}^n \rightarrow \mathbb{R},
\end{equation}
where \(\mathbf{x}\in\mathcal{X}\) denotes a vector of inputs, here GA hyperparameters, and \(f(\mathbf{x})\) is the maximum specific elastic modulus achieved during the corresponding GA optimization run.
Each evaluation of \(f\) is stochastic, as GA involves randomness in the crossover and mutation stages (see Section \ref{GA}), while obtaining even one noisy observation can take several days (complete GA run with FFT). BO addresses this by building a probabilistic surrogate model and iteratively selecting new candidate points that balance exploration and exploitation using an acquisition function. At each BO iteration, a new set of hyperparameters is proposed, evaluated through $q$ repeated GA runs to account for stochastic variability, and incorporated into the surrogate model as noisy observations. The surrogate is then updated using the accumulated data, progressively improving the estimation of promising regions in the hyperparameter space, while minimizing the number of expensive black-box evaluations. 

\subsubsection{Gaussian process surrogate (low fidelity)}\label{sec:GP}

A GP is employed in the BO framework as a probabilistic surrogate model of the expensive black-box objective function. In the present work, the GP approximates the relationship between the GA hyperparameters and the resulting maximum specific elastic modulus obtained during optimization. Since GP evaluations are computationally inexpensive compared to full GA optimization runs assisted by FFT simulations, the GP constitutes the low-fidelity level of the proposed multi-fidelity framework.

Within the proposed BO framework, the GP surrogate is updated after each iteration using the maximum specific elastic modulus $E$ obtained for a given set of GA hyperparameters. Besides predicting promising hyperparameter configurations, the GP provides an uncertainty estimate that identifies regions of the search space that remain insufficiently explored.

In a GP, the kernel (covariance) function defines the correlation between points in the input space and therefore controls the smoothness and characteristic length scales of the surrogate model. Several kernel functions, including isotropic and anisotropic as well as stationary and non-stationary kernels, have been proposed and successfully applied across a wide range of Gaussian Process modeling problems \citep[e.g., ][]{rasmussen2006gaussian,noack_autonomous_2020,hernandez-del-valle_robotically_2023,noack_unifying_2024,Ozdemiretal,schenk2026noiseaware}. In this work, an anisotropic Matérn 5/2 kernel was employed as it permits moderately irregular behavior while retaining sufficient smoothness for efficient optimization.

This Matérn 5/2 kernel is defined as
\begin{equation}
k(\mathbf{x}_i,\mathbf{x}_j)=
\sigma_f^2
\left(
1+\sqrt{5}\;r_{ij} +\dfrac{5}{3}\;r_{ij}^2\right)
\exp\left(
-\sqrt{5}\;r_{ij}
\right),
\end{equation}
where $r_{ij}=\sqrt{\sum_{k=1}^d\frac{(x_{i,k}-x_{j,k})^2}{\ell_k^2}}$ is the Automatic Relevance Determination (ARD)-scaled distance between the two input vectors,
\(\ell_k>0\) is the characteristic lengthscale associated with the \(k\)-th input dimension,
and $\sigma_f^2$ is the signal variance. Following the BoTorch implementation of \emph{get\_matern\_kernel\_with\_gamma\_prior} from \emph{the gpytorch\_modules}, Gamma priors were placed on the kernel lengthscales and output scale during hyperparameter estimation. 

\subsubsection{Acquisition functions}
Alongside the GP surrogate, the acquisition function constitutes the second key component of BO, guiding the search by balancing exploration of uncertain regions and exploitation of promising solutions. An acquisition function maps the GP posterior, characterized by its predictive mean $\mu(x)$ and standard deviation $\sigma(x)$, to a scalar value used to select the next hyperparameter configuration for evaluation. Because evaluations of the objective function are noisy and computationally expensive, the choice of acquisition function plays a critical role in guiding the BO process efficiently. Numerous acquisition functions have been proposed in the literature \citep{shahriari_taking_2016,frazier2018tutorial,letham2019constrained,ament2023unexpected}. In the present study, four acquisition functions were investigated and compared.

Upper Confidence Bound (UCB) is a simple, analytically cheap acquisition function containing explicit terms to balance exploitation and exploration:

\begin{equation}
\alpha(\mathbf{x};\lambda) = \mu(\mathbf{x}) + \lambda\,\sigma(\mathbf{x}),
\end{equation}

where \(\lambda>0\) controls this exploration-exploitation balance. Larger values of $\lambda$ increase the contribution of uncertainty and encourage exploration of poorly sampled regions of the search space. 

Another widely used acquisition function is Expected Improvement (EI) which quantifies the expected improvement relative to the best objective value observed so far \(f(\mathbf{x}^\star)\). 

Defining the improvement random variable as

\begin{equation}
    I(\mathbf{x})=\max\bigl(0,\,f(\mathbf{x})-f(\mathbf{x}^\star)\bigr),
\end{equation}
the expected improvement is given by
\begin{equation}
    \mathrm{EI}(x)=\mathbb{E}[I(\mathbf{x})],
\end{equation}
where the expectation is taken with respect to the GP posterior distribution. EI therefore favors points that are predicted to outperform the current best solution while also accounting for predictive uncertainty.

When observations are noisy, the best observed value \(f(\mathbf{x}^\star)\) is itself uncertain, which can limit the effectiveness of standard EI. To address this issue, Noisy Expected Improvement (NEI) integrates over the posterior uncertainty in both the latent objective function and the observations, typically using Monte Carlo sampling. The batch variant, qNEI, further extends this formulation by jointly selecting a set of \(q\) candidate points and evaluating their expected improvement collectively.

The logarithmic variant, logNEI, applies a logarithmic transformation to the expected improvement, improving the numerical conditioning of the acquisition function when improvement values become very small \citep{ament2023unexpected}. This transformation often leads to more robust optimization behavior and more reliable optimization of the acquisition function itself. Compared with NEI, logNEI places greater emphasis on relative gains and can reduce the tendency to overexploit regions associated with highly uncertain predictions.

\section{Results and discussion}
This section first describes the BO setup used for hyperparameter optimization, followed by a comparison of acquisition functions, the resulting optimal hyperparameters, and the effects of introducing the penalized objective.
\subsection{Genetic algorithm and Bayesian optimization setup}

The BO run was initialized by generating an initial set of 25 Sobol points covering the three-dimensional hyperparameter space: the number of parents to produce offspring for the next generation $n_{par} \in [10,\ 175]$, the fraction of offspring selected for mutation $f_{mut} \in [0.0,\ 1.0]$, and the fraction of cells to mutate in an offspring $f_{cell} \in [0.0,\ 0.75]$. To facilitate BO, all hyperparameters were first normalized to the interval $[0,1]$. Sampling then was performed in the normalized space, and the resulting values were mapped back to their physical ranges before each GA run. For example, a normalized value of $\tilde{n}_{par}=0.4$ corresponds to $n_{par}=76$ in the GA.

\begin{figure}[h]
    \centering
    \includegraphics [width=0.5\textwidth]{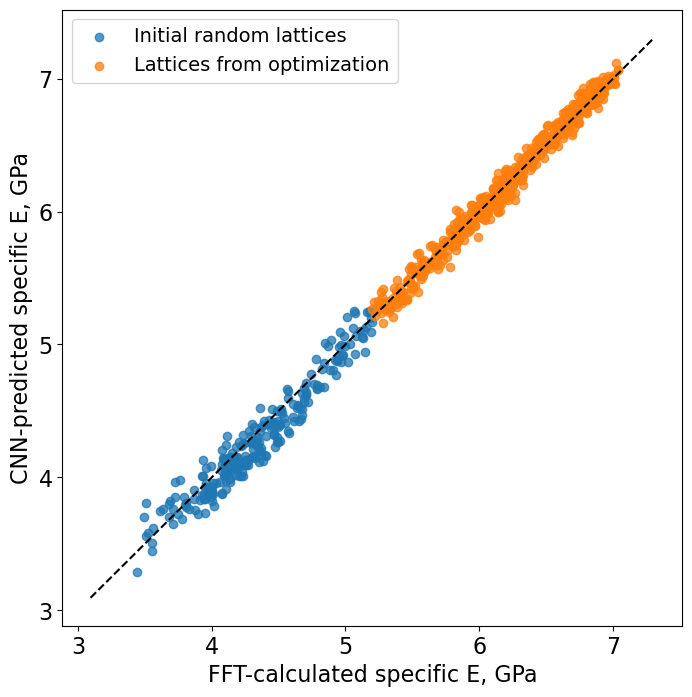}
    \caption{Calculated versus predicted specific elastic modulus values for the test subset.}\label{general_model}
\end{figure}

There is an important difference in the GA setup between this optimization study and the framework described in \citet{zorkaltsev2026does}. Previously, FFT simulations were a key part of active learning, enabling the model to adjust its weights according to newly generated lattices. As BO requires many evaluations of a given black-box function, it would be computationally inefficient to run FFT. Therefore, during BO, all GA evaluations were carried out using the CNN surrogate, while full FFT-assisted GA runs were reserved exclusively for validation of the final hyperparameter configurations identified by BO. To enable this, the same DenseNet architecture was retrained with an expanded data set comprising all simulated lattices generated over 75 generations of single-direction optimization (Section 3.3 in \citet{zorkaltsev2026does}). Figure~\ref{general_model} compares the specific elastic modulus predicted by the CNN surrogate with the corresponding values obtained from FFT simulations. Achieving an MAE of 0.063 GPa and an RMSE of 0.081 GPa, the model effectively steers the GA optimization by selecting the most promising candidates with the highest specific $E$, thereby eliminating the need for FFT evaluations during every iteration. The entire BO pipeline was implemented with the BoTorch library \citep{balandat2020botorch}, and the 3DCNN surrogate was built, trained, and tested with PyTorch \citep{pytorch}.

\subsection{Optimized hyperparameters}
Four acquisition functions were evaluated within the same BO framework: UCB ($\beta = 1.0$), NEI, qNEI ($q = 3$), and logNEI. For each acquisition function, BO was run for 30 iterations using the same set of 25 initial Sobol points. In all cases, the GA was run for 25 generations with an identical initial population.

\begin{figure}[h]
    \centering
    \includegraphics [width=0.95\textwidth]{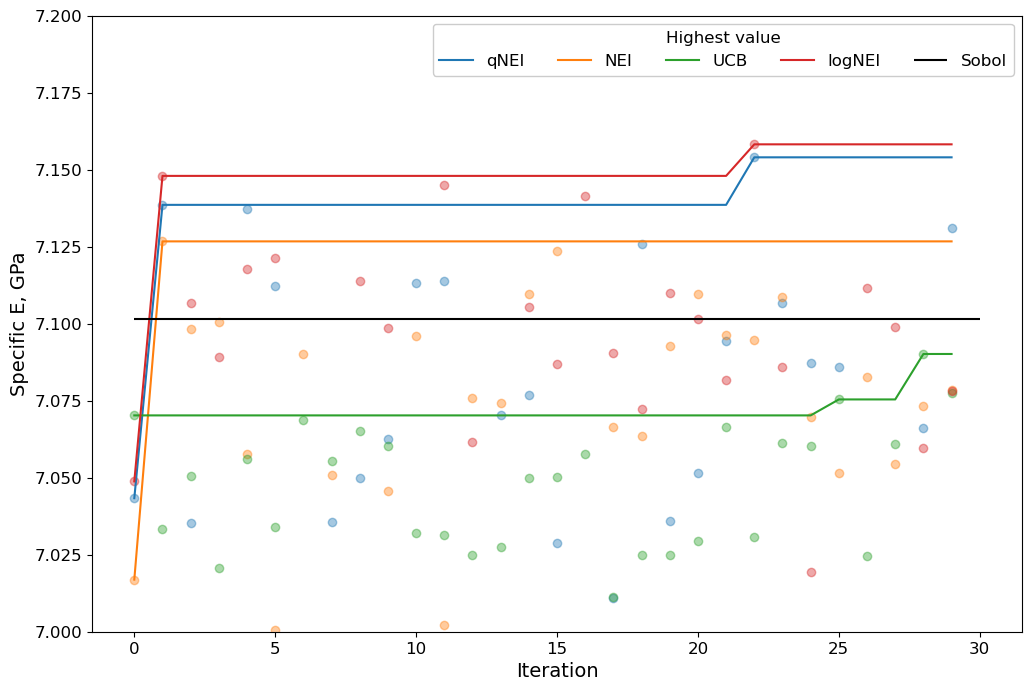}
    \caption{Evolution of the best specific elastic modulus during BO for different acquisition functions (UCB, NEI, qNEI, and logNEI). All optimizations were initialized with the same 25 Sobol points and run for 30 iterations.}\label{acq_functions}
\end{figure}

Figure~\ref{acq_functions} illustrates the evolution of the best specific elastic modulus obtained during the hyperparameter optimization process. Among the acquisition functions tested, UCB exhibited the weakest performance, failing to improve the best candidate identified in the initial Sobol design. NEI achieved better results by explicitly accounting for noise in GA evaluations, while qNEI further enhanced performance through batch sampling of candidate hyperparameter configurations. The highest objective values were obtained with logNEI, which applies a logarithmic transformation to the objective function. Although most acquisition functions converged toward regions associated with high objective values, qNEI and logNEI explored near the global optimum more effectively, demonstrating a superior balance between exploration and exploitation. The hyperparameter configurations corresponding to the best-performing solutions are summarized in Table \ref{acq_summary}.

\begin{table}
\centering
\caption{Optimal hyperparameter configurations identified by BO using different acquisition functions compared to best Sobol initial sampling point.}\label{tab1}%
\begin{tabular}{@{}lcccc@{}}
\toprule
Acq.\ function & $E_{max}$, GPa  & $n_{par}$ & $f_{mut}$ & $f_{cell}$ \\
\midrule
UCB	     & 7.090  & 149  & 0.116 & 0.435 \\
NEI	     & 7.127  & 175  & 0.000 & 0.715 \\
qNEI	 & 7.154  & 153  & 0.184 & 0.000 \\
logNEI	 & 7.158  & 165  & 0.646 & 0.044 \\
\midrule
Sobol	& 7.102  & 170  & 0.234 & 0.287  \\

\bottomrule
\end{tabular}
\label{acq_summary}
\end{table}

A particularly notable result is that the hyperparameters identified by logNEI allowed GA to achieve a specific elastic modulus of 7.158 GPa after only 25 generations, closely matching the value of 7.119 GPa obtained after 75 generations in the complete optimization reported in \citet{zorkaltsev2026does}. This highlights the practical benefit of BO, which reduced the computational cost by 24\% while maintaining the same level of objective performance. The reported computational costs correspond to the GA optimization runs performed on identical hardware and do not include the one-time cost of training the CNN surrogate model.

Furthermore, evaluating the hyperparameters of the original, non-optimized GA framework (($n_{par}=100$), ($f_{mut}=1.0$) and ($f_{cell}=0.0625$)) using the BO surrogate trained during logNEI optimization yielded a predicted specific elastic modulus of 6.37 GPa. This prediction agrees well with the value achieved after 25 generations in the single-direction optimization study (6.51 GPa) reported in \citet{zorkaltsev2026does}. The close correspondence between the predicted and observed values confirms that the surrogate model effectively guided the BO process and provides reliable estimates of GA performance within the explored hyperparameter space.

The comparison of acquisition functions was performed using a fixed optimization setup and a single BO trajectory for each acquisition function. While the observed trends provide useful insights into their relative performance, future work should investigate the robustness of these findings across multiple optimization runs and random seeds.

\subsection{Effects of objective with penalization term}

The results, demonstrated in the previous section, showed that BO can effectively fine-tune the parameters of GA to maximize the provided objective, while also reducing the number of generations required for convergence. However, there are practical limitations arising when moving from simulation-based to experiment-based optimization. Some of the hyperparameter combinations identified as optimal require a large number of parents, \(n_{par}\), to be selected at each iteration. While this remains computationally efficient in the fully CNN-based framework, it quickly becomes problematic when each of the generated offspring needs to be 3D printed and mechanically characterized.

To address this issue, we modified the objective function by introducing a penalty for large parent populations. The penalized objective was defined as:

\begin{equation}
    F_{pen}(x) = \frac{F(x)}{1 + \alpha \times n_{par}},
    \label{eq:pen}
\end{equation}

where \(F(x)\) is the maximum specific $E$ achieved in 25 generations of GA, \(\tilde{n}_{par} \in [0,\ 1]\) is the normalized number of structures selected as parents in each generation, and \(\alpha\) is a penalization parameter. BO with this modified objective shifts the target from the absolute highest achieved specific $E$ to the trade-off between the highest value and the number of structures. 

BO with the penalized objective was performed four times using different penalization weights \(\alpha=\ 0.10,\ 0.15,\ 0.20,\ 0.25\). All four optimization runs used the logNEI acquisition function, which achieved the best performance among the acquisition functions evaluated in the previous section. Table~\ref{norm_summary} summarizes the resulting optimal hyperparameter configurations. The penalized objective values were calculated using Eq~\ref{eq:pen} and served as the objective function during Bayesian optimization, while the corresponding values of the absolute specific elastic modulus are reported for comparison. As expected, increasing \(\alpha\), progressively reduces the number of parents in the optimal hyperparameter configuration. Importantly, this occurs with only a modest reduction in the absolute specific elastic modulus, indicating that there are more efficient GA configurations in the parameter space.
\begin{table}
\centering
\caption{Optimal hyperparameter configurations and corresponding absolute and penalized $E_{max}$ values obtained for different values of the penalization weight $\alpha$.}
\begin{tabular}{@{}lccccc@{}}
\toprule
Value of $\alpha$ & Abs. $E_{max}$, GPa  & Pen. $E_{max}$, GPa  & $n_{par}$ & $f_{mut}$ & $f_{cell}$ \\
\midrule
0.00         & 7.158 & 7.158 & 165  & 0.646 & 0.044 \\
0.10	     & 7.029 & 6.702 & 90  & 0.079 & 0.747 \\
0.15	     & 6.776 & 6.590 & 40  & 0.702 & 0.302 \\
0.20	     & 6.996 & 6.596 & 60  & 0.309 & 0.000  \\
0.25	     & 6.995 & 6.593 & 50  & 0.716 & 0.000  \\
\bottomrule
\end{tabular}
\label{norm_summary}
\end{table}

Without any penalty, optimization has no limits in exploring the provided range of hyperparameters, producing a naturally expected outcome: more parents are selected, more offspring are produced, and there is a higher chance of a better-performing lattice. The penalized objective enabled a nearly twofold reduction in the number of parents without compromising optimization performance. This result further suggests that, although population size plays an important role in GA performance, more efficient locally optimal regions exist within the hyperparameter space and may remain undiscovered when optimizing solely for the original objective function.

\begin{figure}[h]
    \centering
    \includegraphics [width=1.0\textwidth]{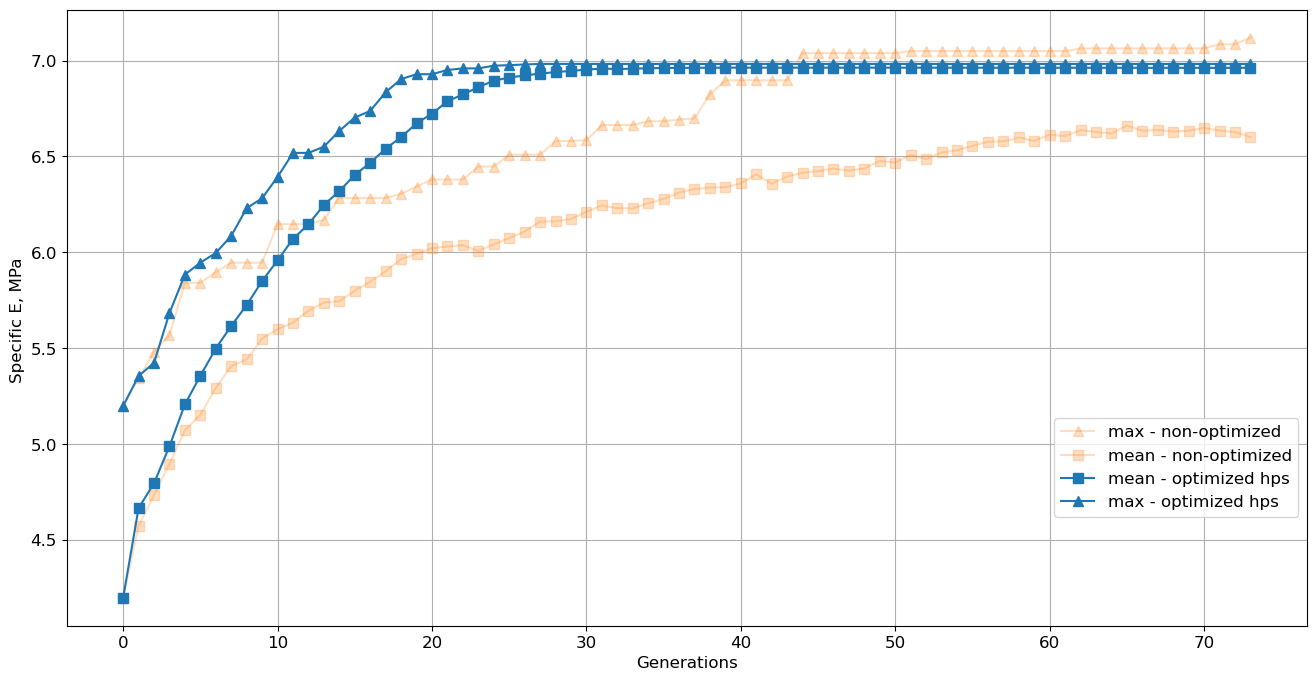}
    \caption{Maximum and mean values of specific elastic modulus in each generation with optimized and non-optimized hyperparameters.}\label{hps_validation}
\end{figure}

The results of hyperparameter validation (Table \ref{norm_summary}, $\alpha =0.20$) are presented in Figure \ref{hps_validation}; the results achieved with the previous 'non-optimized' hyperparameters from \citet{zorkaltsev2026does} are plotted as a reference. The non-optimized run (100 parents and 6.25\% of cells mutated in every offspring) converged to a higher maximum specific modulus (7.119 GPa), and the best-achieved value increased gradually throughout the optimization run. However, the optimized hyperparameter configuration (penalized objective, 60 parents without mutation during the validation run) reached a plateau faster (6.982 GPa), showing no further improvement. The absolute difference between values is 0.137 GPa, or 1.92\%, while the algorithm's behavior was quite different. The optimized configuration reaches a sub-optimal solution and remains constant, and the non-optimized one continues to explore and reaches a slightly higher value.

Removing the mutations completely and reducing the pool of parental lattices drives exploitation, leading to rapid convergence. This strategy is most attractive when experimental and fabrication costs are considered (i.e., the number of lattices to be printed and tested). Employing a larger parent population together with continuous mutation increases the exploratory capability of the GA, which can lead to the identification of better optima. However, this benefit is offset by the need to simulate or fabricate a larger number of lattice structures, thereby increasing the overall optimization time. For instance, the hyperparameters optimized with a penalized objective resulted in 2.28 hours/generation during the GA, while the initial 'non-optimized' hyperparameters required 3 hours per generation on the same hardware.

The disappearance of mutation in the optimized configuration is particularly noteworthy. For the present lattice-design problem, crossover between high-performing parent structures appears sufficient to generate competitive offspring, reducing the benefit of additional random perturbations. This suggests that the search space contains exploitable structural building blocks that can be effectively recombined without relying heavily on mutation-driven exploration.

From an experimental perspective, reducing the number of parent lattices directly translates into fewer structures that must be fabricated, tested, and analyzed during the optimization process. Consequently, the proposed penalized objective provides a practical mechanism for balancing mechanical performance against experimental cost and time requirements.

\section{Conclusion}
This work demonstrates the effectiveness of a multi-fidelity optimization framework that combines high-fidelity FFT simulations, a DenseNet surrogate model, and Bayesian optimization for efficient genetic algorithm hyperparameter tuning. By replacing most computationally expensive FFT evaluations with surrogate predictions, the framework enabled rapid exploration of the hyperparameter space while retaining sufficient accuracy to identify high-performing configurations.

Among the tested acquisition functions, logNEI produced the best-performing hyperparameters, while NEI and qNEI also outperformed both UCB and the Sobol baseline points. This finding is consistent with the expectation that acquisition functions explicitly accounting for observation uncertainty are well suited to noisy optimization problems such as GA hyperparameter tuning. The optimal hyperparameter configuration identified by BO enabled a 25-generation GA run to achieve specific elastic modulus values comparable to those obtained after 75 generations in the reference optimization study.

Introducing a penalization term for the number of parent lattices revealed a practical trade-off between maximizing stiffness and minimizing the number of required structures. Moderate penalization substantially reduced the parent population size while causing only a minor decrease in the achieved specific elastic modulus. This result is particularly relevant for experimental applications, where the fabrication and testing of additional structures incur significant costs.

Validation using full GA runs with FFT evaluations confirmed the trends predicted by the surrogate-assisted optimization. The optimized hyperparameter sets achieved rapid convergence and maintained nearly identical mechanical performance while requiring fewer parent lattices and lower computational effort. Overall, Bayesian optimization reduced the computational cost by approximately 24\% (from 225 h to 171 h) while preserving most of the mechanical performance, demonstrating its value as a practical framework for efficient hyperparameter tuning in lattice optimization problems.

There are several directions in which the proposed framework can be further extended. First, experimental validation through 3D printing and compression tests could introduce an additional level of fidelity beyond those considered in the present study. Second, the optimized hyperparameters identified here are expected to be most reliable for lattice systems, objective functions, and surrogate models similar to those considered in this work. Assessing their transferability to substantially different optimization problems remains an important topic for future investigation.

\backmatter

\section*{Acknowledgments and Funding Information}
This publication is part of the R\&D\&I project PCI2022-132975, funded by MICIU/AEI/10.13039/501100011033, and by the European Union NextGenerationEU/PRTR.

Additionally, C.S. acknowledges funding through a Ramón y Cajal grant (Grant No. RYC2024-048744-I) awarded by the Spanish Ministry of Science and Innovation and financed by MICIU/AEI/10.13039/501100011033 and FSE+.

\section*{Conflict of interest}
The authors report that they have no competing interests to declare.

\section*{Author contribution}
S.Z.: Writing - original draft, Writing - review and editing, Methodology, Software, Validation, Investigation, Formal analysis.
C.S.: Writing - original draft, Writing - review and editing, Methodology, Software, Validation, Investigation, Formal analysis, Supervision, Conceptualization, Funding acquisition.
M.H.: Writing - original draft, Writing - review and editing, Methodology, Conceptualization, Supervision, Project administration, Funding acquisition.
All authors have read and agreed to the current version of the manuscript.

\section*{Replication of results}
Data and code are available in a public repository on GitHub at \href{https://github.com/sergei-zor/ga-bayesian-opt}{https://github.com/sergei-zor/ga-bayesian-opt}.


\begin{appendices}






\end{appendices}


\bibliography{sn-bibliography}

\end{document}